\documentclass[preprint,prc,showpacs,preprintnumbers,amsmath,amssymb,floatfix]{revtex4-2}
\usepackage{amsmath}
\usepackage{graphicx}% Include figure files
\usepackage{dcolumn}% Align table columns on decimal point
\usepackage{bm}% bold math
\usepackage{ulem} %% for strike-through
\usepackage[usenames]{color}
\usepackage{epstopdf}
\usepackage{epsfig}
\usepackage{float}
\usepackage{subfigure}
\usepackage{multirow}
\usepackage[colorlinks,linkcolor=blue,citecolor=red]{hyperref}

\newcommand{\beq}{\begin{equation}}
	\newcommand{\eeq}{\end{equation}}
\newcommand{\beqa}{\begin{eqnarray}}
	\newcommand{\eeqa}{\end{eqnarray}}
\newcommand{\be}{\begin{eqnarray}}
	\newcommand{\ee}{\end{eqnarray}}

\begin{document}

\title{Resolving the spurious-state problem in Dirac equation by using the staggered-grid method}% Force line breaks with \\

\author{Lingfeng Li}
\affiliation{School of Physics, Nankai University, Tianjin 300071, China}

\author{Jinniu Hu}%
\email{hujinniu@nankai.edu.cn}
\affiliation{%
	School of Physics, Nankai University, Tianjin 300071, China\\
	and Shenzhen Research Institute of Nankai University, Shenzhen 518083, China
}

 \author{Ying Zhang}
 \email{yzhang@tju.edu.cn}
\affiliation{Department of Physics, School of Science, Tianjin University, Tianjin 300072, China}

\author{Hong Shen}
\affiliation{School of Physics, Nankai University, Tianjin 300071,  China}

\date{\today}% It is always \today, today,

\begin{abstract}
	Discretizing the Dirac equation on a uniform grid with the central difference formula often generates spurious states. We propose a staggered-grid scheme in the framework of the finite-difference method that suppresses these spurious states without introducing Wilson terms or ad-hoc filtering. In this approach, the large and small components of the Dirac equation are placed on interlaced nodes, and the first-order derivatives are evaluated between staggered points, yielding a Hamiltonian that breaks the unitary transformation between $H_\kappa$ and $H_{-\kappa}$. Benchmarks with the nuclear Woods–Saxon potentials demonstrate one-to-one agreement with the eigenvalues {obtained from shooting method and asymmetric finite-difference method, rapid convergence for weakly bound states, and reduced box-size sensitivity}. The method retains the simplicity of central differences and standard matrix diagonalization, while naturally extending  to higher-order and multi-dimension systems. It provides a compact and efficient tool for relativistic bound-state and scattering calculations.

\end{abstract}

\maketitle

\section{Introduction\label{sec1}}

The Dirac equation occupies a central place in quantum theory because it unifies nonrelativistic quantum mechanics with special relativity and provides a first-principles description of spin-$1/2$ fermions \cite{thaller2013dirac,bagrov2014dirac}. In its covariant form,
\begin{equation}
	(i\gamma^\mu \partial_\mu - m)\,\psi(x)=0,
\end{equation}
it explains the intrinsic spin and magnetic moment of the electron, predicts the existence of antiparticles, and yields the fine-structure and spin–orbit splitting observed in atomic spectra. Beyond atomic physics, the Dirac framework was also applied by modern nuclear mean-field models for nucleons with the strong scalar–vector fields \cite{walecka1974theory}, the quark-level description of hadrons \cite{nohl1975bound}, and effective quasirelativistic theories in condensed-matter systems (e.g., graphene, Dirac and Weyl semimetals) \cite{castro2009electronic}. It is also essential for describing particles in extreme fields and densities relevant to heavy elements, superheavy candidates, laser–plasma interactions, and compact-star matter.

Many physically practical problems involving Dirac Hamiltonians do not own analytical solutions. Numerical solvers are therefore indispensable for studying bound states, resonances embedded in the continuum, and scattering observables. Reliable discretization must preserve Hermiticity and the symmetries of the continuum operator, avoid spectral pollution and variational collapse, and handle both the $r\to 0$ and $r\to\infty$ behaviours in radial problems. In spherical symmetry, for example, the coupled first-order radial equations for the large and small components demand accurate first derivatives and stable coupling across a wide range of energies, including weakly bound and near-continuum states. In the past one hundred years, many well-established computational methods have been applied to solve the eigenvalue problems of Dirac equation, such as the shooting method, Green's function theory, imaginary time step scheme, and so on \cite{horowitz1981self,meng2006relativistic,sun2020green,zhang2010avoid}.

Another way is discretizing the derivative term of the kinetic operator in uniform grid and diagonalizing the matrix of Hamiltonian \cite{li2024one}. However it may generate the unphysical eigenpairs—spurious-state problem from central differences and spectral pollution from the unbounded spectrum, i.e. “fermion doubling” problem, where the wave functions of spurious states are highly oscillating.  In lattice gauge theory, a Wilson term lifts the doubler branch but breaks chiral symmetry and introduces an extra scale, while  domain-wall and overlap fermions improve chiral properties with adding algorithmic complexity~\cite{Wilson1974PRD,susskind1977lattice,ukawa2015kenneth,zhao2016spherical}. Variational frameworks that respect the Dirac min–max structure, together with high-order difference formulas, can guarantee pollution-free spectra but typically involve heavier implementation and mesh management~\cite{dolbeault2000eigenvalues,salomonson1989relativistic,fang2020solution}. For self-consistent fields, inverse-Hamiltonian or imaginary-time evolution schemes can avert variational collapse but give up the convenience of one-shot matrix diagonalization~\cite{hagino2010iterative,tanimura20153d,ren2017solving,li2020efficient}.

Recently, we successfully developed an asymmetric (parity-alternating) finite-difference (AFD) method, that can completely remove the spurious states of Dirac equation while retaining finite-difference simplicity \cite{zhang2022resolving} by using the asymmetric difference formula (ADF). The behaviours of  wave functions at large distance are less dependent on the box size.  It has also been extended to solve the relativistic Hartree-Bogoliubov equation and reasonably describe the properties of exotic neutron-rich nuclei  \cite{wang2025solving}.

Staggered-grid discretization has a long history in computational physics and engineering \cite{hustedt2004mixed,liu2014optimal}. The central idea is to place different field components (or different components of the solution) on interlaced locations—cell centers, faces, edges, or half steps—so that discrete gradient–divergence and curl–curl pairs inherit the algebraic dualities of the continuum operator, {which reflects the high numerical stability of this discretization}. Classic examples include the Marker-and-Cell (MAC) scheme in incompressible fluid dynamics \cite{harlow1965numerical}, which uses velocity–pressure staggering to eliminate checkerboard pressure modes and strengthen the coupling between momentum and continuity, and the Yee scheme in computational electromagnetic, which staggers electric and magnetic fields in space and time to preserve Faraday’s and Ampère–Maxwell laws with low numerical dispersion \cite{yee1966numerical}.  Compared with collocated grids, staggered layouts naturally suppress odd–even decoupling, reduce spurious modes, and pair cleanly with high-order differences, finite-volume fluxes, and energy-conserving time integrators. A closely related idea in lattice field theory is the Kogut–Susskind fermion, which distributes spinor degrees of freedom over sublattices to mitigate fermion doubling \cite{KogutSusskind1975PRD}. 

In this work, we will apply the staggered-grid finite-difference method (SGM) to solve the eigenvalue problem of Dirac equation with Woods–Saxon potentials. By placing the large and small components on interlaced nodes and evaluating first derivatives between staggered points, the unitary transformation symmetry of the lattice Hamiltonians for $H_\kappa$ and $H_{-\kappa}$ is broken.  The spurious states in the conventional central difference formula are expected to be eliminated. The results will be compared to those from shooting and AFD methods. 

This paper is written as follows. In Section II, we introduce the basic framework of the SGM to solve the Dirac equation. In Section III, the numerical results of Woods-Saxon potentials are shown and discussed. In Section IV, a summary and perspective will be given.

\section{Staggered-grid finite-difference method\label{sec2}}
At first, we consider a particle with mass $M$ and in the scalar $S(\bm{r}))$ and vector $V(\bm{r})$ potentials. The corresponding Dirac equation is given by
\begin{eqnarray}
	\{\bm{\alpha\cdot p}+V(\bm{r})+\beta[M+S(\bm{r})]\}\Psi(\bm{r})=\varepsilon\Psi(\bm{r}),
\end{eqnarray}
where $\bm{\alpha}$ and $\beta$ are Dirac matrices, and $\varepsilon$, $\Psi(\bm{r})$ represent the eigenvalues and eigenvectors, respectively.

In a static spherical system, the wave function of the particle can be expressed as
\begin{eqnarray}
	\Psi(\bm{r})=\frac{1}{r}\begin{pmatrix}
		G(r)Y_{ljm}\\
		iF(r)Y_{ljm}
	\end{pmatrix},
\end{eqnarray}
where $Y_{ljm}$ are spin spherical harmonics, $j=l\pm1/2$, and $G(r)$, $F(r)$ are the large and small radial components of $	\Psi(\bm{r})$, respectively. The radial equation about $G$ and $F$ is written as
\begin{eqnarray}
	\begin{pmatrix}
		\Sigma(r)&\frac{\kappa}{r}-\frac{d}{dr}\\
		\frac{\kappa}{r}+\frac{d}{dr}&\Delta(r)
	\end{pmatrix}\begin{pmatrix}
	G(r)\\F(r)
	\end{pmatrix}=E\begin{pmatrix}
	G(r)\\F(r)
	\end{pmatrix},
	\label{eq3}
\end{eqnarray}
where $\Sigma$ and $\Delta$ are combinations of scalar and vector potentials,
\begin{eqnarray}
	\begin{aligned}
		\Sigma(r)&=V(r)+S(r),\\
		\Delta(r)&=V(r)-S(r)-2M,\\
		E&=\varepsilon-M,\\
		\kappa&=(-1)^{j+l+1/2}(j+1/2).
	\end{aligned}
\end{eqnarray}

We would like to take the nucleus $^{132}$Sn $(N=82,~Z=50)$ as an example to solve the Dirac equation.  A Woods-Saxon form potential is adopted to describe the finite nuclei for the $\Sigma(r)$ and $\Delta(r)$ fields. Specifically, their expressions are
\begin{eqnarray}
	\begin{split}
		V(r)+S(r)&=\frac{V^0}{1+e^{\frac{r-R_0}{a}}},\\
		V(r)-S(r)&=\frac{-\lambda V^0}{1+e^{\frac{r-R^{ls}}{a^{ls}}}},
	\end{split}
	\label{eq.1}
\end{eqnarray}
where $R_0=r_0A^{1/3}$, $R^{ls}=r^{ls}A^{1/3}$. For the neutron, the central potential depth is modified as
\begin{eqnarray}
	V^0=V\left(1-\kappa\frac{N-Z}{N+Z}\right).
	\label{eq.2}
\end{eqnarray}
The physical parameters in Eqs.~(\ref{eq.1}) and (\ref{eq.2}) were obtained by fitting the neutron results from relativistic mean-field model as shown in Table \ref{tab:2} from Ref.~\cite{Koepf1991}.

\begin{table}[htbp]
	\caption{\label{tab:2}  Parameters of Woods-Saxon potential.}
	\begin{tabular}{ccccccc}
		\hline\hline
		~$V$ [MeV]~ &~ $\lambda$ ~&~ $\kappa$ ~&~ $r_0$ [fm] ~& $r^{ls}$ [fm] ~& $a$ [fm] ~&~ $a^{ls}$ [fm] \\ 
		\hline
		$-71.28$ & $11.12$ & $0.462$ & $1.233$ & $1.144$ & $0.615$ & $0.648$  \\ 
		\hline\hline
		\end{tabular}
	\end{table}

When we attempt to apply the FDM to solve the Dirac equation, wave functions of large and small components would be discretized into a set of points along with the radial coordinate at a fixed interval $h$ from $0$ to the box size $R_\text{box}=nh$. The first-order differential operator in Eq.~(\ref{eq3})  is replaced by the finite-difference formula. A common choice is using the central difference formula (CDF) as from $h$ to $nh$
\begin{eqnarray}
	\frac{df(r)}{dr}\approx\frac{f(r+h)-f(r-h)}{2h}.
	\label{fdf}
\end{eqnarray}

The large and small components $G(r)$ and $F(r)$ are transformed into a column vector in the FDM
\begin{eqnarray}
	\begin{pmatrix}
		G(r)\\F(r)
	\end{pmatrix}=\begin{pmatrix}
	G(h)\\G(2h)\\\vdots\\G(nh)\\F(h)\\F(2h)\\\vdots\\F(nh)
	\end{pmatrix}.
	\label{eq.8}
\end{eqnarray}

Therefore, the radial Dirac equation, Eq.~(\ref{eq3}) could be rewrite as the matrix form
\begin{eqnarray}
	\begin{split}
		\Sigma(ih)G(ih)+\frac{\kappa}{ih}F(ih)-\frac{F(ih+h)-F(ih-h)}{2h}=EG(ih),\\
		\frac{\kappa}{ih}G(ih)+\frac{G(ih+h)-G(ih-h)}{2h}+\Delta(ih)F(ih)=EF(ih),
	\end{split}
	\label{eq.9}
\end{eqnarray}
where $i=1,~2~\cdots~n$. Eq.~(\ref{eq.9}) in the discretized coordinate space can be expressed as a block matrix equation, 
\begin{eqnarray}
	\begin{pmatrix}
		\begin{array}{c|c}
			A & B_1\\
			\hline
			B_2 & C\\  
		\end{array}
	\end{pmatrix}\begin{pmatrix}
	G\\F
	\end{pmatrix}=E\begin{pmatrix}
	G\\F
	\end{pmatrix}.
\end{eqnarray}

Among them, the matrix corresponding to the first-order derivative operator with CDF should be
\begin{eqnarray}
	\frac{d}{dr}=\frac{1}{2h}\begin{pmatrix}
		0&1&&&&\\
		-1&0&1&&&\\
		&&\ddots&\ddots&&\\
		&&&-1&0&1\\
		&&&&-1&0
	\end{pmatrix}.
\end{eqnarray}
Accordingly, $G(0),~F(0),~G((n+1)h)$ and $F((n+1)h)$ are taken to be zero in Eq.~(\ref{eq.9}), resulting in an $n\times n$ matrix for the first-order derivative operator. Here, $B_1,~B_2$ are tridiagonal matrices, while $A,~C$ are diagonal. The resulting $2n \times 2n$ Hamiltonian matrix is Hermitian. By diagonalizing this Hamiltonian matrix numerically, we can get $2n$ sets of eigenvalues and corresponding wave functions.

However, solutions obtained by the three-point CDF (3PCDF) exhibit pairs of degenerate solutions between $\kappa=1$ and $\kappa=-1$, known as spurious states, characterized by the rapidly oscillating wave functions~ \cite{zhang2022resolving}. This arises because the derivative of a point by using CDF depends only on the wave function values at the adjacent points, not at the point itself. For instance, the data sequences like $1,~3,~1,~3,~1$ and $1,~1,~1,~1,~1$ yield the same central difference values according to Eq.~(\ref{fdf}). The coupling between $F$ and $G$ in Eq.~(\ref{eq3}) is effectively weakened in Eq.~(\ref{eq.9}) due to this reason, resulting in partial decoupling—known in computational fluid dynamics as checkerboard oscillation~\cite{harlow1965numerical}. Mathematically, the reason of the generation of degenerate physical and spurious states is the existence of a unitary transformation $U$ which could transform the Hamiltonian $H_{\kappa}$ into $H_{-\kappa}$ through $UH_{\kappa}U^{-1}=H_{-\kappa}$ \cite{zhao2016spherical}.

To reduce the number of spurious states, higher-order schemes like five-point CDF have been proposed \cite{zhang2022resolving}. However, they do not eliminate the spurious state completely due to  the lack of direct coupling between components. As an alternative, we applied the ADF instead of CDF to solve the Dirac equation, which successfully resolved the spurious-state problem completely. 

\begin{figure}[h]
	\includegraphics[width=0.85\linewidth]{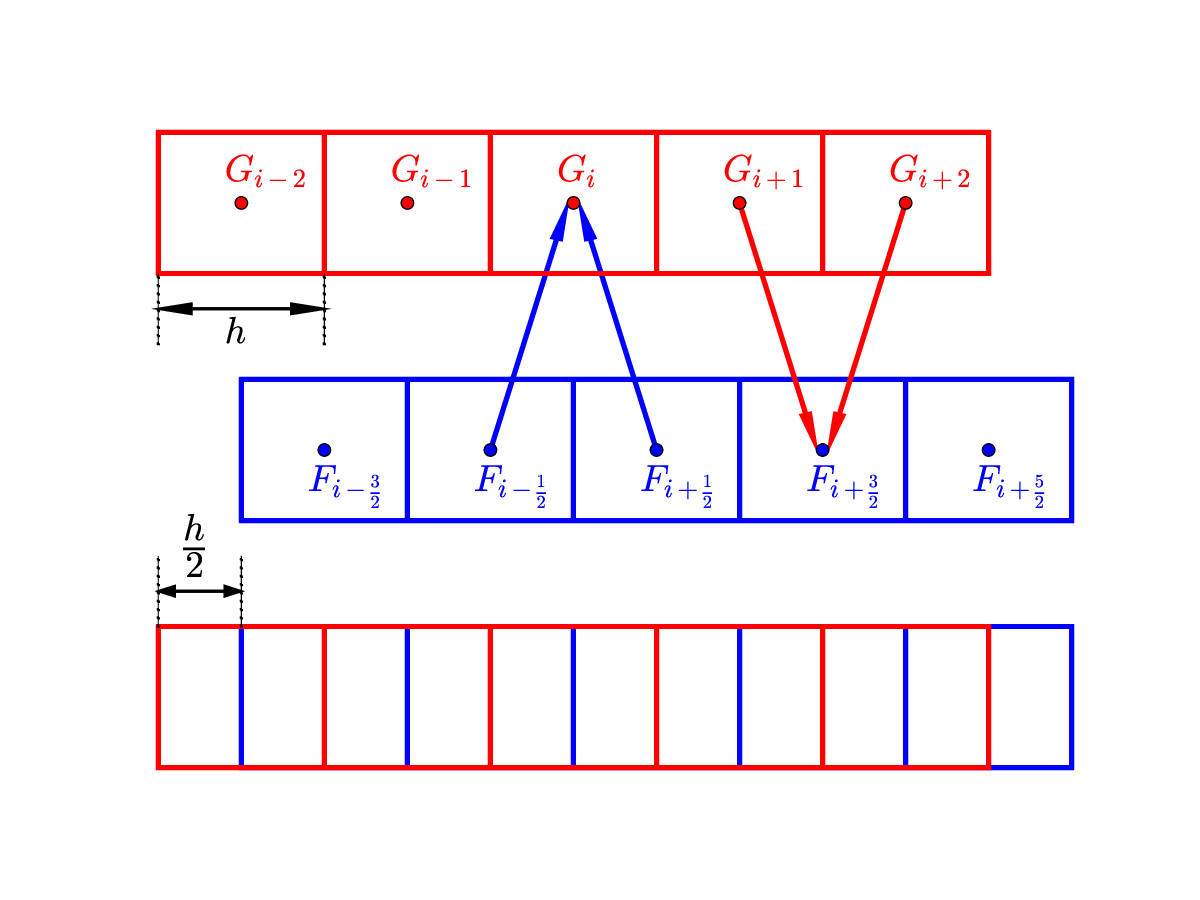}% Here is how to import EPS art
	\caption{\label{fig:1} In the Schematic of SGM,  large and small components, $G(r)$ and $F(r)$ are shifted by $h/2$.}
\end{figure}

On the other hand, the SGM, inspired by the MAC method for incompressible fluid flow~\cite{harlow1965numerical}, offers another solution to avoid the oscillation in solving the normal differential equations. Unlike the wave functions on the same grid  in Eq.~(\ref{eq.8}), the positions of large and small components of Dirac equation $G$ and $F$, shift with $h/2$ in SGM, as illustrated in Fig.~\ref{fig:1}.

Therefore, the corresponding column vector of wave functions has the form,
\begin{eqnarray}
	\begin{pmatrix}
		G(r)\\F(r)
	\end{pmatrix}=\begin{pmatrix}
	G(h)\\G(2h)\\\vdots\\G(nh)\\F(1/2h)\\F(3/2h)\\\vdots\\F((n-1/2)h)
	\end{pmatrix}.
	\label{eq.12}
\end{eqnarray}

Now, the first-order derivate of the wave function at $r$ can be calculated by the wave functions at $r-1/2h$ and $r+1/2h$
\begin{eqnarray}
	\frac{df(r)}{dr}\approx\frac{f(r+1/2h)-f(r-1/2h)}{h},
	\label{eq.13}
\end{eqnarray}
which replaces the forward or backward differential formula in AFD method \cite{zhang2022resolving}, and retains second-order accuracy.

The first-order derivative terms in Eq.~(\ref{eq.9}) can be easily shifted and returned to the matrix equation based on Eq.~(\ref{eq.12}). Applying this to Eq.~(\ref{eq.9}), one must use an expression of the centrifugal term $k/r$ at shifted grids according to the wave functions. Actually, we found that, by using a linear interpolation, these terms can keep the matrix equation Hermiticity, which can be written as
\begin{eqnarray}
	&&\frac{1}{ih}F(ih)\approx\Big[\dfrac{1}{(2i-1)h}F(ih-1/2h)+\dfrac{1}{(2i+1)h}F(ih+1/2h)\Big],
	\nonumber\\
	&&\frac{1}{(i-1/2)h}G(ih-1/2h)\approx\dfrac{1}{(2i-1)h}[G((i-1)h)+G(ih)].
	\label{eq.15}
\end{eqnarray}

Under the above approximation Eq. \eqref{eq.15}, the $B_1$ and $B_2$ matrices can be written as
\begin{widetext}
\begin{eqnarray}
	\renewcommand{\arraystretch}{1.5}
	B_1=\begin{pmatrix}
		\frac{1}{h}+\frac{\kappa}{h}&-\frac{1}{h}+\frac{\kappa}{3h}&&&&\\
		&\frac{1}{h}+\frac{\kappa}{3h}&-\frac{1}{h}+\frac{\kappa}{5h}&&&\\
		&&&\ddots&&\\
		&&&&\frac{1}{h}+\frac{\kappa}{(2N-3)h}&-\frac{1}{h}+\frac{\kappa}{(2N-1)h}\\
		&&&&&\frac{1}{h}+\frac{\kappa}{(2N-1)h}\\
	\end{pmatrix},
\end{eqnarray}
\begin{eqnarray}
	\renewcommand{\arraystretch}{1.5}
	B_2=\begin{pmatrix}
		\frac{1}{h}+\frac{\kappa}{h}&&&&\\
		-\frac{1}{h}+\frac{\kappa}{3h}&\frac{1}{h}+\frac{\kappa}{3h}&&&\\
		&\ddots&\ddots&&\\
		&&-\frac{1}{h}+\frac{\kappa}{(2N-3)h}&\frac{1}{h}+\frac{\kappa}{(2N-3)h}&\\
		&&&-\frac{1}{h}+\frac{\kappa}{(2N-1)h}&\frac{1}{h}+\frac{\kappa}{(2N-1)h}\\
	\end{pmatrix},
\end{eqnarray}
\end{widetext}
with $B_1=B_2^T$. 

In this way, we can diagonalize the Hamiltonian matrix constructed by SGM to obtain the eigenvalues and wave functions of Dirac equation. Unlike AFD method, it does not require to alternately adopt the backward or forward difference formula for different $\kappa$ values.

\section{\label{sec3}Numerical results and discussions}
The numerical calculations to solve Dirac equation with Woods-Saxon potential were performed in a spherical box with a radius of $R_{\text{box}}=20$ fm, using $n=500$ lattices. The convergence with respect to the lattice number has been clearly checked. The results are well converged for $n=500$ lattices.  The bound-state energies at the $ns_{1/2}(\kappa=-1)$ and $np_{1/2}(\kappa=1)$ neutron states in $^{132}$Sn,  obtained using three-point SGM (3PSGM) are shown in Table \ref{tab:1}. For comparison, the results obtained by the three-point ADF (3PADF) are also listed in the table.  The energy of $1s_{1/2}$ state from 3PSGM is $-55.006$ MeV, whereas $-55.004$ MeV from 3PADF and $-55.005$ MeV from shooting method. For the higher-energy states, such as $2s_{1/2}$, the energy from SGM is $-33.933$ MeV, which is closer to the result from shooting method, $-33.929$ MeV, comparing to that from 3PADF, $-33.915$ MeV. For the other states, the calculations from  3PSGM have higher accuracies in contrast to those from AFD method.

\begin{table}[htbp]
	\caption{\label{tab:1}The neutron energy levels of $^{132}$Sn for $\kappa=-1$ and $\kappa=1$ in the Woods-Saxon potential calculated by SGM and AFD method with different difference formulas. The unit of the energy is MeV.}
	\begin{ruledtabular}
		\begin{tabular}{cccccc}
			3PSGM & 3PNHSGM & 3PADF & 5PSGM & 5PNHSGM & 5PADF \\ 
			\hline
			$\kappa=-1$&&&&&\\
			$-55.006$ & $-55.006$ & $-55.004$ & $-55.005$ & $-55.003$ & $-55.005$ \\
			$-33.933$ & $-33.932$ & $-33.915$ & $-33.931$ & $-33.924$ & $-33.929$ \\
			$-9.219$  & $-9.217$  & $-9.171$  & $-9.213$  & $-9.204$  & $-9.210$ \\
			$\kappa=1$&&&&&\\
			$-46.162$ & $-46.163$ & $-46.157$ & $-46.162$ & $-46.162$ & $-46.162$ \\
			$-21.407$ & $-21.409$ & $-21.377$ & $-21.404$ & $-21.405$ & $-21.405$ \\
			$-0.306$  & $-0.307$  & $-0.283$  & $-0.304$  & $-0.305$  & $-0.304$ \\
		\end{tabular}
	\end{ruledtabular}
\end{table}

To obtain more accurate results, the five-point difference formula is adopted in SGM as the 5PSGM, whose results are also given in Table \ref{tab:1}. They are also compared to energy levels from the corresponding 5-point ADF (5PADF) as we discussed in the previous work \cite{zhang2022resolving}. It should be noted that, the results given by the shooting method are exactly the same with those obtained by the 5PADF, which is not shown in the table.  It can be found that the calculations of 5PSGM are closer to those from 5PADF and shooting method.

We also investigate another approximation method to treat the centrifugal term $\kappa/r$ in the first equation of Eq.~(\ref{eq.15}) related to $F$, which is
\begin{eqnarray}
	\dfrac{1}{ih}F(ih)\approx\dfrac{1}{ih}\dfrac{1}{2}[F(ih-1/2h)+F(ih+1/2h)]
\end{eqnarray}
while the second equation of Eq.~(\ref{eq.15}) about $G(r)$ remains the same. This form maintains the symmetry of the equations and reduces approximation errors in the denominator, although it leads to a non-Hermitian Hamiltonian matrix. Remarkably, its diagonalization still yields real eigenvalues, which have been seriously examined with different numerical methods and codes. They are shown in Table~\ref{tab:1} as the three-point non-Hermitian SGM (3PNHSGM). They have similar accuracies with those from 3PSGM. Moreover, the corresponding wave functions from 3PNHSGM  are identical to those obtained by 3PSGM. Similarly, we also extended this approximation with five-point difference formula as 5PNHSGM, whose results are consistent with those from 5PSGM.  

\begin{figure}[htbp]
	\includegraphics[width=0.6\linewidth]{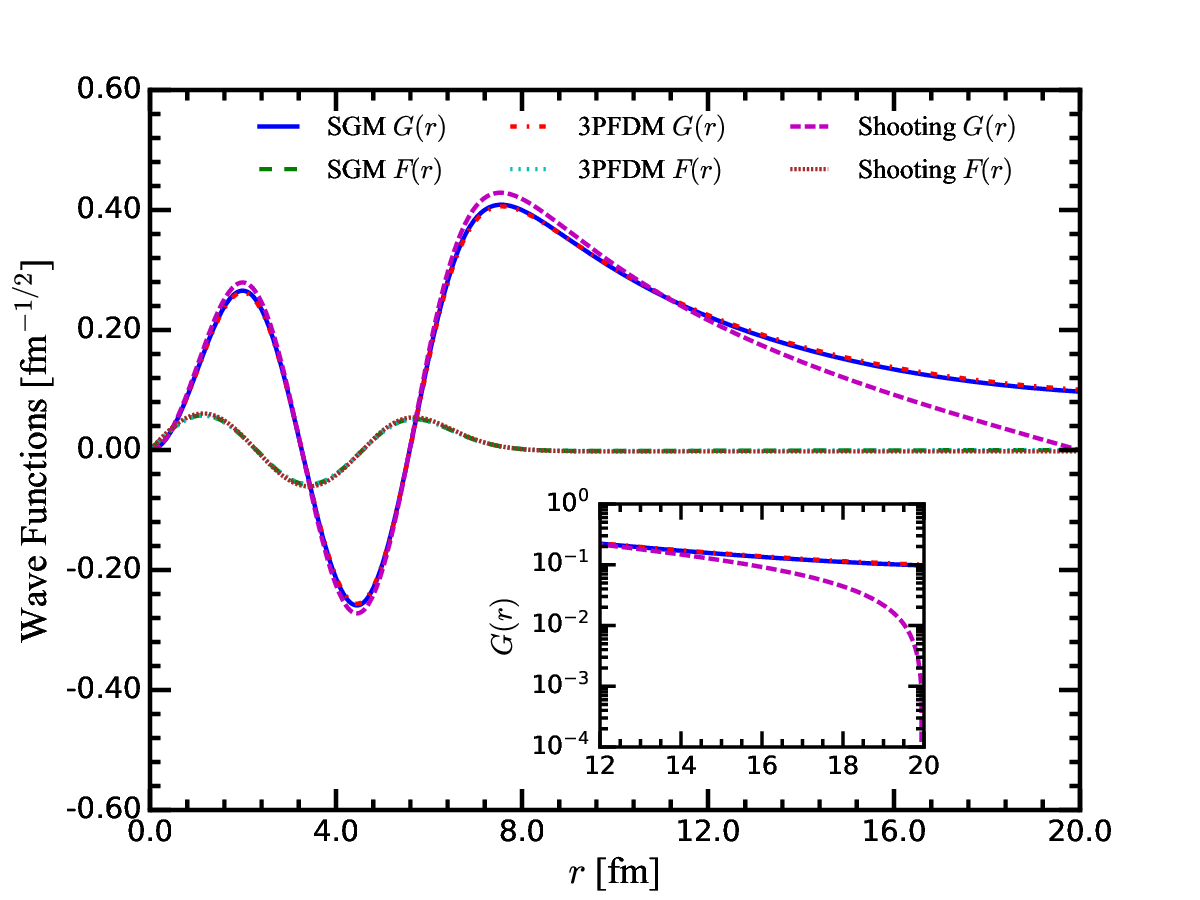}% Here is how to import EPS art
	\caption{\label{fig:2} The wave functions $G(r)$ and $F(r)$ of $3p_{1/2}$ in $^{132}$Sn  from the 3PSGM, 3PFDM, and shooting method.}
\end{figure}

The large and small components, $G(r)$ and $F(r)$ of $3p_{1/2}$ state from 3PSGM, 3PFDM, and shooting method are shown in Fig.~\ref{fig:2}, respectively. For the deeper bound states with $\kappa=\pm 1$, the wave functions from above three methods are completely identical. Meanwhile, $3p_{1/2}$ state is a weak bound state, with $E=-0.306$ MeV from 3PSGM. In principle, the large component $G$ should have a long tail at large distance to exhibit its asymptotic behaviour. However, the shooting method usually assumes that the $G$ should be zero at the box size for the bound states. Therefore, $G(r)$ will quickly decline to zero at box boundary in shooting method.  On the other hand, in 3PSGM and 3PFDM, there is no restriction of the wave function at the boundary $r=R_\text{box}$, while the wave function outside $r=R_\text{box}$ is assumed to be zero. The wave functions from these two methods naturally satisfy the requirement of asymptotic behaviour, where $G(R_\text{box})$ is not zero. The asymptotic tails of the wave functions are clearly given in the insert panel of  Fig.~\ref{fig:2}.  For the small components, $F(r)$, the results from three methods are almost the same.

\section{\label{sec4}Conclusion}

In summary, the Dirac equation in spherical system with Woods-Saxon potential has been successfully solved by using the SGM with central difference formula (CDF). {By discretizing the large and small components of the wave function on staggered grids by a half-step shift, the usual checkerboard oscillation problem in first-order derivative calculation by CDF, i.e. spurious-state problem is perfectly eliminated}. {This phenomenon demonstrates the high numerical stability of the SGM method}.

To satisfy the requirements of staggered grid in wave function,  the centrifugal terms, $\kappa/r$ in Dirac equation at lattice should be treated with some approximations. The first one can preserve the Hermiticity of Hamiltonian matrix with 3-point CDF, which is inferred as 3PSGM. The other will slightly break the Hermiticity of Hamiltonian matrix, called as 3PNHSGM. However, we found that the real eigenvalues can still be obtained from 3PNHSGM, which have been carefully inspected. The results of 3PSGM and 3PNHSGM are comparable to those from the AFD method with 5-point difference formula. The asymptotic behaviour of weak bound states at large distances are well ensured.

In this work, we proposed a novel numerical method to solve the spurious-state problem of Dirac equation when its first-order derivative is discretized by CDF. {The SGM, due to its simplicity, efficiency, and accuracy, could be a valuable tool for solving the Dirac equation}, which can be applied to the investigation of quantum dots, nuclear many-body system, and so on, or fewer symmetry constraint object.

\section{Acknowledgments}
This work was supported  by the National Natural Science Foundation of China Nos. 12175109, 12475149, and the Natural Science Foundation of Guangdong Province (Grant  No: 2024A1515010911). 

\section{Data Availability}
The  python code and data that support the findings of this article are openly available in zenodo.

\bibliographystyle{apsrev4-2}
\bibliography{ref}% Produces the bibliography via BibTeX.

\end{document}